\documentclass[10pt,a4paper]{article}
\usepackage{makeidx}
\usepackage{amssymb}
\usepackage{amsmath}
\usepackage{graphicx}
\usepackage{graphics}
\usepackage{times}
\usepackage[T1]{fontenc}
\usepackage[thinspace,thinqspace,squaren]{SIunits}

\usepackage{ctable}
\allowdisplaybreaks

\usepackage{algorithm}
\usepackage{algorithmic}

\twocolumn

\begin{document}
\date{}

\title{Autonomous take-off and
landing of a tethered aircraft:\\ a simulation study}
\author{Eric Nguyen Van, Lorenzo Fagiano and Stephan Schnez
\thanks{The authors are with ABB Switzerland Ltd., Corporate Research, 5405 Baden-D\"{a}ttwil - Switzerland}
\thanks{E-mail addresses: \{ eric.nguyen-van $|$ lorenzo.fagiano $|$ stephan.schnez\}@ch.abb.com}}
\maketitle

\begin{abstract}
The problem of autonomous launch and landing of a tethered rigid aircraft for airborne wind energy generation is addressed. The system operates with ground-based power conversion and pumping cycles, where the tether is repeatedly reeled in and out of a winch installed on the ground and linked to an electric motor/generator. In order to accelerate the aircraft to take-off speed, the ground station is augmented with a linear motion system composed by a slide translating on rails and controlled by a second motor. An onboard propeller is used to sustain the forward velocity during the ascend of the aircraft. During landing, a slight tension on the line is kept, while the onboard control surfaces are used to align the aircraft with the rails and to land again on them. A model-based, decentralized control approach is proposed, capable to carry out a full cycle of launch, low-tension flight, and landing again on the rails. The derived controller is tested via numerical simulations with a realistic dynamical model of the system, in presence of different wind speeds and turbulence, and its performance in terms of landing accuracy is assessed. This study is part of a project aimed to experimentally verify the launch and landing approach on a small-scale prototype.
\end{abstract}

\section{Introduction}\label{S:intro}
Airborne wind energy (AWE) systems aim to convert wind energy
into electricity by exploiting tethered aircrafts, whose motion is stabilized by active control systems \cite{AWEbook,FaMi12}. The advantages of this concept over traditional wind turbines are the lower construction and installation costs and the possibility to reach stronger winds blowing at  higher altitudes, in the range of 400-600 m above ground. The main challenge is the high complexity of the system. 
Currently, AWE is an umbrella-name for a set of different specific implementation approaches which can
be classified by the way the lift force 
is generated - either aerodynamic lift 
\cite{ch20-ZiHa15,ch21-MiTM14,ch23-vdVlPS14,ch24-BRKGS14,ch26-RuSo14,ch28-Vand14},
or aerostatic lift 
\cite{ch30-VeGR14} - and by the placement of the
electrical generators - either on-board of the aircraft
\cite{ch28-Vand14,ch30-VeGR14} or on the ground
\cite{ch20-ZiHa15,ch21-MiTM14,ch24-BRKGS14,ch23-vdVlPS14,ch26-RuSo14}.
Among the systems that exploit aerodynamic lift and ground-level
generators, a further distinction can be made between concepts that
rely on rigid wings \cite{ch26-RuSo14}, similar to gliders, and
concepts that employ flexible wings like power kites
\cite{ch20-ZiHa15,ch21-MiTM14,ch23-vdVlPS14,ch24-BRKGS14}.
Small-scale prototypes (10-50 kW of rated power) of the
mentioned concepts have been realized and successfully tested to
demonstrate their power generation functionalities. Moreover,
scientific contributions concerned with technical
aspects like aerodynamics
\cite{BSTR14,ch16-BrSO14,ch17-BSTR14,ch18-GoLu14,ch19-LRBLJP14},
controls
\cite{HoDi07,CaFM09c,BaOc12,ErSt12,FZMK14,FHBK14,FeSc14,ch12-ZaGD14,ErSt14,ZaFM15},
resource assessment \cite{ArCa09,ch5-Arch15}, economics
\cite{FaMP09,ch7-ZiHa15}, prototype design \cite{FaMa15}, and
power conversion \cite{SHVDD15}, have recently appeared, gradually
improving and expanding our understanding of such systems.

Notwithstanding the continuous development of the field, some relevant aspects still need to be addressed in order to ultimately
prove the commercial feasibility of the idea. One of such aspects is
the autonomous launch and landing of the aircraft with a relatively small ground area occupation, particularly for concepts that employ
rigid wings and ground-level power conversion. In fact, while systems
with on-board generation \cite{ch28-Vand14,ch30-VeGR14} and kite-based
systems with ground generation \cite{ch20-ZiHa15} are able to take-off autonomously from a quite compact ground
area, the same functionality for AWE systems with rigid wings and
ground-level generators has not been fully demonstrated yet. Indeed, there is
evidence of autonomous take-off of this class of generators
\cite{ampyx}, however by using a launching procedure that requires
a significant space, virtually in all directions in
order to adapt to the wind conditions during take-off. As a
consequence, one of the main advantages of AWE systems, i.e. the
possibility of being installed at low costs in a large variety of
locations, 
might be lost due to the need of a large area of land
suitable for the take-off. So far, only few studies in the literature address this problem, and all of them consider only the launching phase. In \cite{ZaGD13}, a rotational start-up of the aircraft is studied and simulated, with a focus on the control and optimization aspects of the approach. In \cite{Bont10}, an analysis of several launching techniques is carried out, considering different performance criteria, and three alternatives are deemed the most promising: buoyant systems, linear ground acceleration plus on-board propeller, and a rotational start-up. The rotational start-up is then examined in more detail by means of numerical simulations.

At ABB Corporate Research, we recently started to actively address this problem via theoretical, numerical and experimental research. Our goal is to understand the tradeoffs between land occupation, aircraft characteristics, and electrical power required to carry-out the launch and landing phases, and ultimately to assess their impact on the overall technical and economical feasibility of the energy generation concept.
To this end, we consider a rigid tethered glider, conceptually similar to the ones employed by the company Ampyx Power \cite{ampyx,ch26-RuSo14}, and we study a linear launch technique where the ground station is augmented with a linear motion system that accelerates the aircraft to take-off speed. 
In this paper, we present part of our results, in particular the derivation of a mathematical model of the system including the ground station and the tethered glider, and the design of a model-based control system able to carry out autonomously a cycle of launching, low-tension flight, and landing. 
We finally present numerical simulations to evaluate the accuracy and repeatability of the landing maneuver in the presence of different nominal wind speed and turbulence. These results were instrumental to design the control system for a small-scale experimental setup, whose testing is currently ongoing.

\section{System description and problem formulation}\label{S:System_description}
\subsection{System description}\label{SS:yo_yo_descr}
\begin{figure}[!htb]
 \begin{center}
  \includegraphics[trim= 0cm 0cm 0cm 0cm,width=8cm]{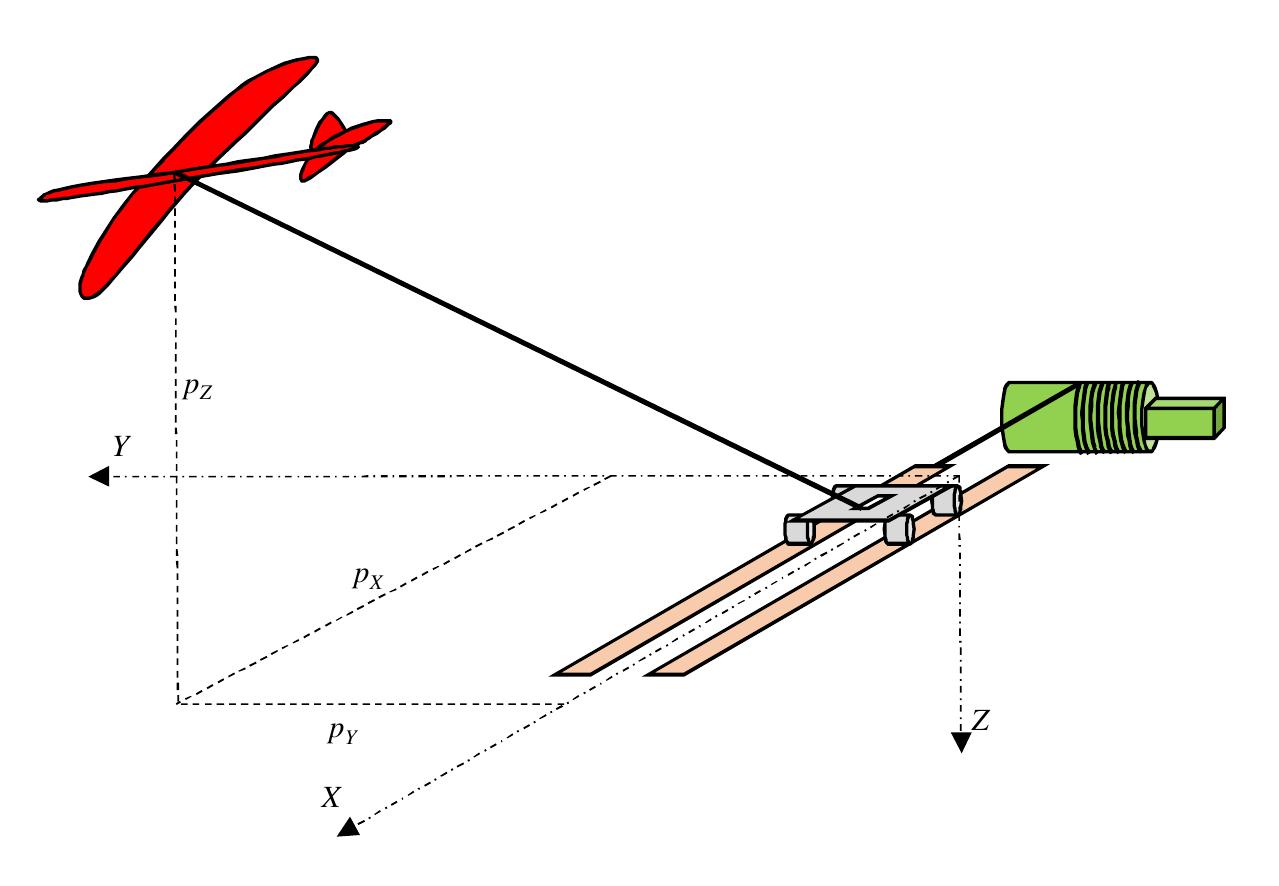}
  \caption{Sketch of the considered system together with the $(X,Y,Z)$ inertial frame and the position $[p_X,\,p_Y,\,p_Z]^T$ of the aircraft.}
  \label{F:sys_sketch}
 \end{center}
\end{figure}

\begin{figure}[!tbh]
 \begin{center}
  \includegraphics[trim= 0cm 0cm 0cm 0cm,width=8cm]{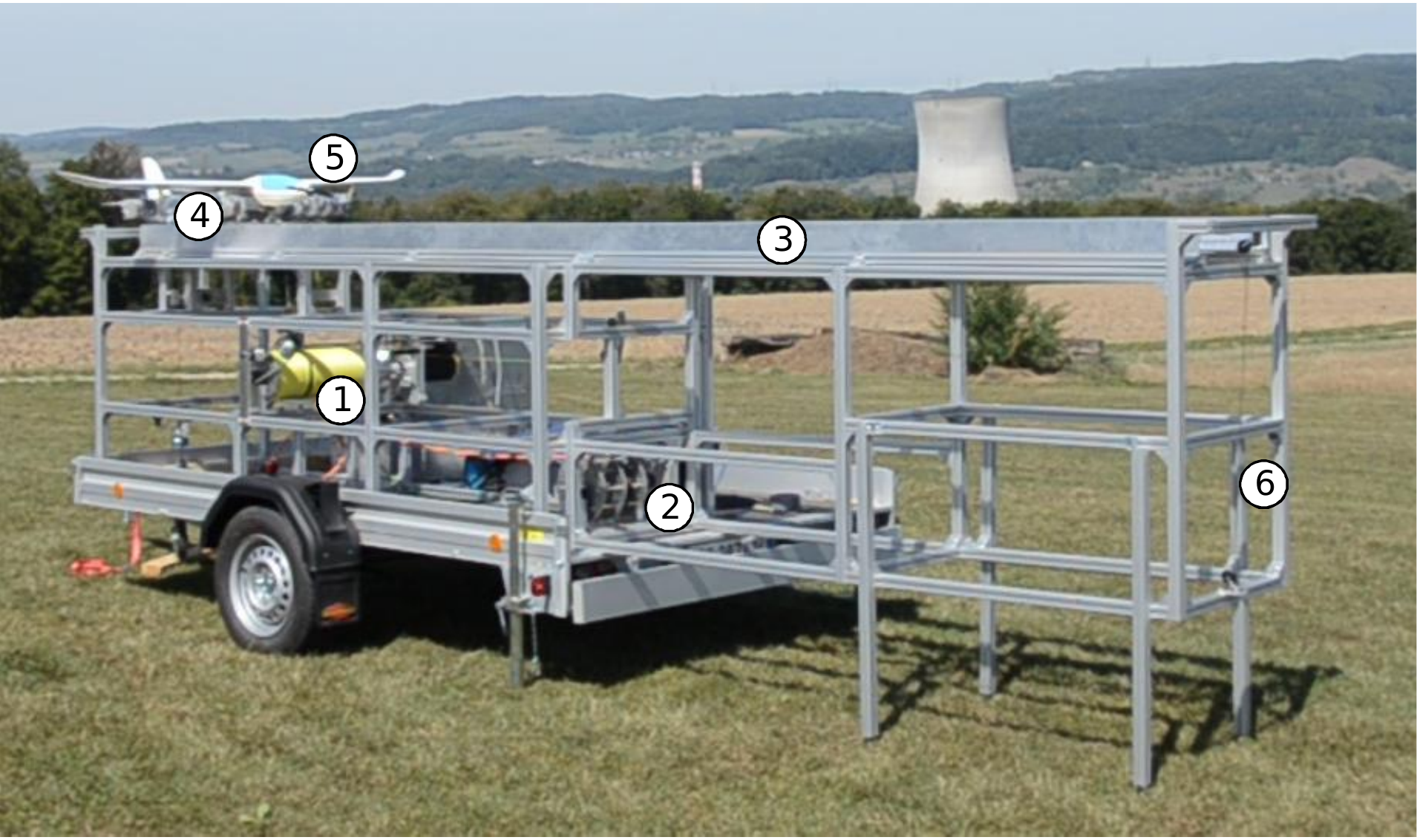}
  \caption{Picture of the small-scale prototype built at ABB Corporate Research. The numbers in the picture indicate: 1. the winch, 2. the motor that moves the slide via a second tether, 3. the rails, 4. the slide, 5. the glider, 6. the aluminum frame.}
  \label{F:sys_pic}
 \end{center}
\end{figure}

The system we consider is composed of a ground station equipped with a winch, storing a tether connected to a rigid glider, see Fig. \ref{F:sys_sketch} for a sketch and Fig. \ref{F:sys_pic} for a picture of our small-scale experimental setup.
Two electric machines are installed on the ground station: the first one controls the winch in order to achieve, during power production, a repetitive cycle of reeling-out under high load, hence converting the mechanical power into electrical, and of reeling-in under low load, spending a small fraction of the energy to start a new production phase. The second machine controls the movement of a linear motion system composed of a slide on rails. 
A series of pulleys redirects the line from the winch to the slide and then to the glider. The position and speed of both electric machines are measured via encoders and hall-effect sensors, the position of the ground station is measured via GPS, finally the tension on the line is measured with a load cell installed at the attachment point of one of the pulleys on the ground station. All these measurements can be used for feedback control of the ground station. The available manipulated variables on the ground are the torques of the two electric machines. The tether is made of ultra-high-molecular-weight polyethylene fibers braided to obtain a high-strength, low density material with 8-15 times larger strength-to-weight ratio than steel.

We consider a glider with a conventional design with a single foldable electric propeller in the front. The glider's attitude, absolute position, angular rates and linear velocity vector are measured with an Inertial Measurement Unit (IMU) and a GPS. The incoming airspeed along the body longitudinal axis is also measured, using an air speed sensor. The available control surfaces are the ailerons (for roll control), elevator (for pitch control), rudder (for yaw control), and flaps (to increase lift and drag during take-off and landing). Together with the front propeller, these form the five manipulated variables available to influence the glider's motion. In our experimental setup (see Fig. \ref{F:sys_pic}) the aircraft is a commercially available model made of foam, which is inexpensive, easy to modify and resilient to impacts.
\subsection{System operation} \label{Operation}
For the sake of the present study, we can divide the desired system operation in three phases:

\textbf{Launching}. The glider is initially attached to the slide. The slide is accelerated to take-off speed and decelerated down to rest within the length of the rails. The glider starts its on-board propulsion during the acceleration and detaches from the slide when the peak speed has been reached. 

\textbf{Flight}. Since power generation is not our objective, transition from normal flight to pumping cycles is not considered here. Instead, the aim is to control the flight at relatively low speed notwithstanding the perturbation induced by the tether, and to prepare for the landing procedure. To this end, a roughly rectangular path is executed before approaching the ground for the landing.

\textbf{Landing}. The landing strategy we consider consists in reeling-in the tether to guide the glider to the rails. When the glider is close enough to the ground station, the winch is stopped and the slide is started, hence reeling-in the remaining part of the tether and engaging again with the glider. Finally, the slide is stopped and the tether is used to keep the glider on it during the braking.

\subsection{Control objectives and problem formulation}\label{Goals}
The overall aim of the research activity is to assess the launch and landing capabilities of the described system. This goal translates into specific control objectives for each phase described in section \ref{Operation}. During the launch, the ground station shall be able to synchronously accelerate the slide and the main winch to allow the glider to take-off with low tether tension. During the flight phase, the onboard control unit shall follow the desired path despite the perturbation of the tether and wind turbulence, while at the same time the ground station control system shall reel-out and reel-in the line such that the tension is minimal but non-zero, such that there is little or no slack line. Finally, in the landing phase the glider shall land within the area covered by the rails, to make it feasible to engage the slide again.
%
The problem we address in this paper is to design a control system able to achieve the mentioned goals. We propose an approach which is decentralized, i.e. there is no active communication between the glider and the ground station. Rather, the coordination between the two control systems, onboard and on the ground, is realized by exploiting the measurement of the tether tension. 

\section{Mathematical model of the system}\label{S:aerodynamics}

\begin{figure}[!htb]
 \begin{center}
  \includegraphics[trim= 0cm 0cm 0cm 0cm,width=8cm]{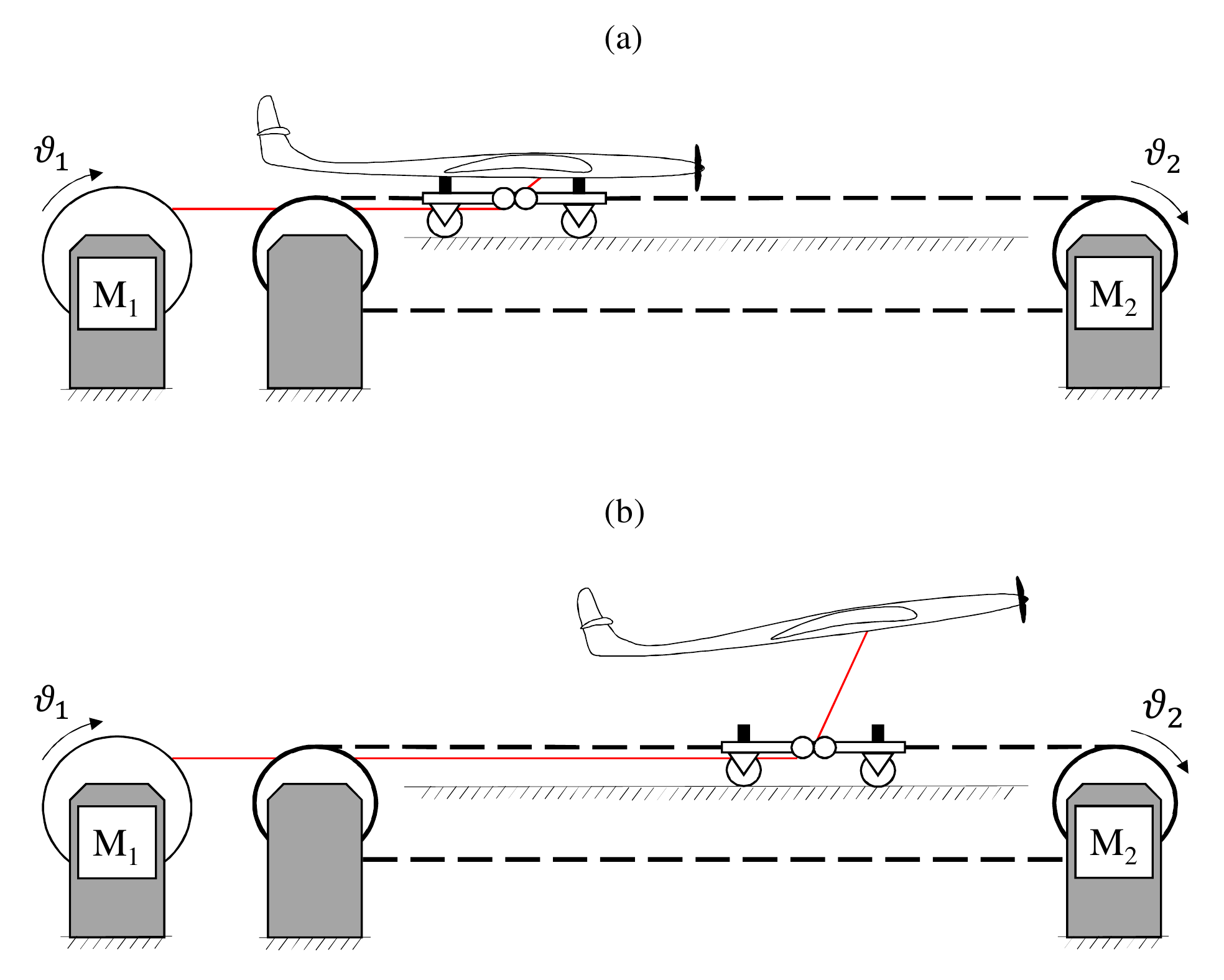}
  \caption{Sketch of the dynamical model. (a) First operating mode, with the aircraft carried by the slide up to take-off speed; (b) second operating mode, with the aircraft detached from the slide.}
  \label{F:Dyn_system_sketch}
 \end{center}
\end{figure}

The considered system can be described by a hybrid dynamical model: a first operating mode (Figure \ref{F:Dyn_system_sketch}(a)) describes the system's behavior from zero speed up to the take-off, when the aircraft and the slide can be considered as a unique rigid body; a second operating mode (Figure \ref{F:Dyn_system_sketch}(b)) describes the aircraft motion after take-off, when it is separated from the slide. 

Throughout this section we consider an inertial, right-handed reference frame $(X,\,Y,\,Z)$ with the origin corresponding to the central point of the rails, which are assumed parallel to the ground, the $X-$axis aligned with them, and the $Z-$axis pointing downwards, see Fig. \ref{F:sys_sketch}. We will denote a generic vector in the three-dimensional space as $\vec{v}$ and, when needed, we will specify the reference frame considered to compute the vector's components with a subscript notation, e.g. $\vec{v}_{(XYZ)}$. Each scalar component of the vector will be also followed by its axis, i.e. $\vec{v}_{(XYZ)}=[v_X,\,v_Y,\,v_Z]^T$. All the equations presented in the following have been derived by applying Newton's second law of motion. For the sake of brevity, we omit the explicit dependence of the model variables on the continuous time $t$.

\subsection{Ground station model}\label{SS:ground_station_model}

We denote with $M_1$ the motor/generator linked to the winch and with $M_2$ the one connected to the slide. We indicate with $\vartheta_{M_1}$ the angular position of motor $M_1$, with $\dot{\vartheta}_{M_1}\doteq\frac{d\vartheta_{M_1}}{dt}$ its angular speed, and with $\vartheta_{M_2},\,\dot{\vartheta}_{M_2}$ the angular position and speed of the motor $M_2$. We further denote with $u_{M_1},\,u_{M_2}$ the torques applied by the motors. The state and input vectors of the ground station model are then given by
\begin{equation}\label{E:gs_state}
\begin{array}{rcl}
x_{GS}&\doteq&[\vartheta_{M_1},\,\dot{\vartheta}_{M_1},\,\vartheta_{M_2},\,\dot{\vartheta}_{M_2}]^T\\
u_{GS}&\doteq&[u_{M_1},\,u_{M_2}]^T\\
\end{array}
\end{equation}
Neglecting the elasticity of the linear motion system, the model is described by the following equations in the first operating mode:
\begin{equation}\label{E:dyn_mod_first}
\begin{array}{rcl}
\ddot{\vartheta}_{M_1}&=&\dfrac{1}{J_{M_1}}(r_{M_1}\,\|\vec{F}_t\|_2-\beta_{M_1}\,\dot{\vartheta}_{M_1}+u_{M_1})\\
\ddot{\vartheta}_{M_2}&=&\dfrac{1}{J_{M_2}+(m_s+m)\,r_{M_2}^2}\,((-r_{M_2}\,\|\vec{F}_t\|_2\\
& &+r_{M_2}\,F_{a,X}-r_{M_2}\,\beta_s\,\dot{\vartheta}_{M_2})\\
& &-\beta_{M_2}\,\dot{\vartheta}_{M_2}(t)+u_{M_2})
\end{array}
\end{equation}
where $t$ is the continuous time variable, $r_{M_1}$ is the radius of the winch (assuming for simplicity that the latter is directly connected to the motor/generator), $r_{M_2}$ the radius of the pulley that links motor $M_2$ to the belt, $J_{M_1},\,J_{M_2}$ the moments of inertia of the winch and of the pulley plus their respective motors, $\beta_{M_1},\,\beta_{M_2}$ their viscous friction coefficients, $m_s$ the mass of the slide, $\beta_s$ the viscous friction coefficient of the belt/slide/rail system, $m$ the mass of the aircraft. $F_{a,X}$ and $\vec{F}_t$ are, respectively, the aerodynamic force components along the $X-$axis developed by the aircraft (mainly its drag) and the tether tension force vector. These two forces are described in more detail in sections \ref{SS:glider_model} and \ref{SS:tether_model}.

The switch between the first and the second operating modes takes place at the time instant $t^*$ defined as:
\begin{equation}\label{E:t_star}
t^*=\min\left(\tau\geq0\,:-F_{a,Z}(\tau)>g\,m\right),
\end{equation}
where $F_{a,Z}$ is the component of the aerodynamic force developed by the aircraft along the $Z-$axis and $g$ is the gravity acceleration. Thus, $t^*$ represents the time instant when the aerodynamic lift force of the aircraft is larger than its weight, hence obtaining a positive vertical acceleration.
In the second operating mode, the first equation in \eqref{E:dyn_mod_first} describing the winch dynamics is still valid, while the second one is replaced by:
\begin{equation}\label{E:dyn_mod_second}
\begin{array}{rcl}
\ddot{\vartheta}_{M_2}&=&\dfrac{1}{J_{M_2}+(m_s)\,r_{M_2}^2}\,((-\beta_s\,r_{M_2}^2-\beta_{M_2})\,\dot{\vartheta}_{M_2}+\\
&&+r_{M_2}\,F_{t,X}+u_{M_2})
\end{array}
\end{equation}
where, due to the glider taking off, the mass of the slide is reduced and the tether force is projected along the rails' direction (compare \eqref{E:dyn_mod_first} with \eqref{E:dyn_mod_second}).

\subsection{Aircraft model}\label{SS:glider_model}

\begin{figure}[hbt]
 	\centerline{
 		\includegraphics[width=8cm,clip]{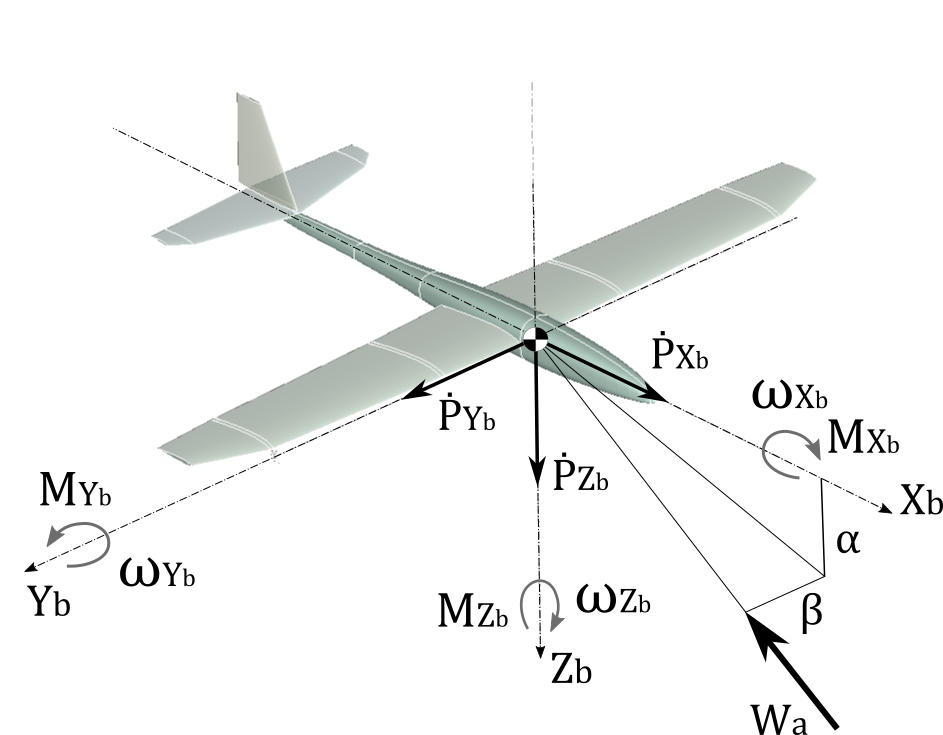}}
 	\caption{Variables and parameters describing the flight expressed in the body frame of reference. $\omega_{X_b}$ is the roll rate, $\omega_{Y_b}$ is the pitch rate and $\omega_{Z_b}$ is the yaw rate. $\alpha$ and $\beta$ are the velocity angles.}\label{D:glider}
 \end{figure}

To introduce the state equations describing the aircraft's dynamics, we consider the body reference frame $(X_b,\,Y_b,\,Z_b)$, represented in Fig.~\ref{D:glider}, which is fixed to the plane and whose rotation relative to the inertial frame $(X,\,Y,\,Z)$ is defined by the Euler angles $\phi$ (roll), $\theta$ (pitch) and $\psi$ (yaw). 
Denoting with $\vec{\omega}$ the angular velocity vector of the glider (see Fig. \ref{D:glider}), the time derivatives of the Euler angles are computed as:
\begin{equation}\label{E:euler_deriv}
\begin{array}{rcl}
\dot{\phi}&= & \omega_{X_b}+(\omega_{Y_b}\sin(\phi) + \omega_{Z_b}\cos(\theta)) \tan(\theta) \\
\dot{\theta}&= & \omega_{Y_b}\cos(\phi) - \omega_{Z_b}\sin(\phi) \\
\dot{\psi}&= & \frac{1}{\cos(\theta)}(\omega_{Y_b} \sin(\phi) + \omega_{Z_b} \cos(\phi))
\end{array}
\end{equation}
We further denote with $\vec{p}$ the position of the aircraft relative to the origin of the inertial system $(X,Y,Z)$. The manipulated variables available for control are denoted with $u_a$ (ailerons), $u_e$ (elevator), $u_r$ (rudder), $u_f$ (flaps), and $u_m$ (motor thrust). In practice, such control surfaces are controlled by servo-motors installed on the aircraft, with position feedback control loops, and each motor is controlled by manipulating the applied voltage. We neglect here the dynamics of such low-level controllers and assume that one can directly manipulate the angular position of the control surfaces and the propeller's trust. This assumption is reasonable for the considered application. We can now define the state and input vectors of the aircraft's model as:
\begin{equation}\label{E:glider_state}
\begin{array}{rcl}
x_g&\doteq&[p_X,\,p_Y,\,p_Z,\,\dot{p}_{X_b},\,\dot{p}_{Y_b},\,\dot{p}_{Z_b},\\
\phi,\,\theta,\,\psi,\,\omega_{X_b},\,\omega_{Y_b},\,\omega_{Z_b}]^T\\
u_g&\doteq&[u_a,\,u_e,\,u_r,\,u_f,\,u_m]^T\\
\end{array}
\end{equation}
as well as the full system's state and input vectors:
\begin{equation}\label{E:full_state}
\begin{array}{rclcrcl}
x&\doteq&\left[\begin{array}{c}x_{GS}\\x_g\end{array}\right]\;\in \mathbb{R}^{16}&\;&u&\doteq&\left[\begin{array}{c}u_{GS}\\u_g\end{array}\right]\in \mathbb{R}^{7}\\
\end{array}
\end{equation}
For a given wind vector $\vec{W}$, the apparent wind velocity $\vec{W}_{a}$ is given by:
\begin{equation}\label{E:apparent_wind}
\vec{W}_{a}=\vec{W}-\vec{\dot{p}},
\end{equation}
i.e. the absolute wind velocity minus the velocity of the aircraft relative to ground.
We further introduce the angle of attack $\alpha$ and the side slip angle $\beta$ of the aircraft (see Fig.~\ref{D:glider}):
\begin{equation}\label{E:alpha_beta}
\begin{array}{rcl}
\alpha&\doteq & \arctan \left(\frac{W_{a,Z_b}}{W_{a,X_b}}\right)\\
\beta&\doteq & \arcsin\left( \frac{W_{a,Y_b}}{\|\vec{W}_a\|_2}\right)
\end{array}
\end{equation}
where $\|\vec{W}_a\|_2$ is the magnitude of the apparent wind velocity. The angles $\alpha,\,\beta$ and their time derivatives $\dot{\alpha},\,\dot{\beta}$ are used to compute the aerodynamic coefficients that, together with $\|\vec{W}_a\|_2$ and the control inputs $u_g$, determine the magnitudes of the aerodynamic force $\vec{F}_a$ and moment $\vec{M}_a$. The orientation of $\vec{F}_a,\;\vec{M}_a$ depends on  the aircraft attitude and on the control inputs, in addition to $\alpha$ and $\beta$. For the sake of space, we omit here the full derivation (see \cite{Etkin} for details). In addition to the aerodynamic effects, we include in the model the thrust of the propeller, $F_{T,X_bY_bZ_b}=[u_m,0,0]^T$, the weight due to the aircraft's mass $\vec{F}_{W,XYZ}=[0,0,m\,g]^T$, and the force $\vec{F}_t$ and moment $\vec{M}_t$ exerted by the line, described in section \ref{SS:tether_model}. The total force and moment applied to the aircraft are computed as $\vec{F}=\vec{F}_a+\vec{F}_{W}+\vec{F}_t+\vec{F}_T$ and $\vec{M}=\vec{M}_a+\vec{M}_t$, respectively, and they are in general a function of the full system's state $x$ and input $u$.
%
%
The following assumptions are considered in the derivation of the aircraft model\cite{Etkin}:
\begin{itemize}
	\item The aircraft is a rigid body with no aero-elastic effects.
	\item The $(X_b,\,Z_b)$ plane of the body axis lies in the symmetry plane of the glider.
	\item For the aerodynamic coefficients, the coupling between the longitudinal and lateral motions is negligible.
	\item The flight takes place at a very low Mach number, thus compression effects are neglected.
\end{itemize}

Under these assumptions, the model equations for the aircraft's dynamics are the following (see Etkin \cite{Etkin}) with variables illustrated in Fig.~\ref{D:glider}:\small
\begin{equation}\label{E:F1}
\begin{array}{rcl}
	\ddot{p}_{X_b}&= {} & \frac{F_{X_b}}{m}  + \omega_{Z_b}\,\dot{p}_{Y_b} - \omega_{Y_b}\,\dot{p}_{Z_b}\\
	\ddot{p}_{Y_b}&= {} & \frac{F_{Y_b}}{m} - \omega_{Z_b}\,\dot{p}_{X_b} + \omega_{X_b}\,\dot{p}_{Z_b}\\
	\ddot{p}_{Z_b}&= {} & \frac{F_{Z_b}}{m} + \omega_{Y_b}\,\dot{p}_{X_b} - \omega_{X_b}\,\dot{p}_{Y_b} \\
	M_{X_b}&= {} & I_{xx}\dot{\omega}_{X_b} - I_{zx}(\dot{\omega}_{Z_b}-\omega_{X_b}\omega_{Y_b}) - (I_{yy}-I_{zz})\omega_{Y_b}\omega_{Z_b} \\
	M_{Y_b}&= {} & I_{yy}\dot{\omega}_{Y_b} - I_{zx}(\omega_{Z_b}^2-\omega_{X_b}^2) -(I_{zz}-I_{xx})\omega_{Z_b}\omega_{X_b} \\
	M_{Z_b}&= {} & I_{zz}\dot{\omega}_{Z_b} - I_{zx}(\dot{\omega}_{X_b}-\omega_{Y_b}\omega_{Z_b})-(I_{xx}-I_{yy})\omega_{X_b}\omega_{Y_b}
\end{array}
\end{equation}\normalsize
where $I_{xx},\ldots,I_{zz}$ denote the components of the inertia matrix of the aircraft computed in the body reference frame.
%
%

\subsection{Tether model} \label{SS:tether_model}
The force $\vec{F}_t$ exerted by the tether is composed of three main contributions: its tension, $\vec{F}_{T,t}$, weight, $\vec{F}_{W,t}$, and drag, $\vec{F}_{D,t}$. 
The tether drag is computed from a moment equilibrium as in e.g. \cite{CaFM09c} and is expressed as:
\begin{equation}\label{E:line_drag}
\vec{F}_{D,t}=\frac{1}{8}\rho C_{t} d_tr_{M_1}\vartheta_{M_1} \|\vec{W}_a\|\vec{W}_a
\end{equation}
Where $\rho$ is the air density, $C_{t}$ the tether drag, and $d_t$ the tether diameter. The tension force is computed using an elastic model where the stiffness is a function of the tether length:
\begin{equation}\label{E:line_tension}
\vec{F}_{T,t}= -\max\left(0,\dfrac{E\,\pi d_t^2}{4\,\varepsilon r_{M_1}\,\vartheta_{M_1}}(\|\vec{p}\|_2-r_{M_1}\,\vartheta_{M_1})\right)\dfrac{\vec{p}}{\|\vec{p}\|_2},
\end{equation}
where $E$ the Young's modulus of the tether and $\varepsilon$ the breaking elongation. Equation \eqref{E:line_tension} highlights the coupling between the ground station dynamics and the aircraft dynamics. We further approximate the weight force applied by the line on the glider as $\vec{F}_{W,t,XYZ}=[0,0,\frac{mg}{2}]$.
%
%
%
%
%
Finally, the moment induced by the tether forces on the glider is expressed as :
\begin{equation}\label{E:line_moment}
\vec{M_t}=
\vec{R} \times \vec{F}_{t}
\end{equation}
where $\vec{R}$ is the vector that points from the aircraft's center of gravity to the attachment point of the tether on the body.

Collecting together equations \eqref{E:gs_state}-\eqref{E:line_moment}, we obtain the overall model as
\begin{equation}\label{E:overall_model}
\dot{x(t)}=f(x(t),u(t),\vec{W}(t)).
\end{equation}
The dynamics are nonlinear, with 16 states, 7 manipulated inputs, and 3 exogenous inputs (i.e. the components of the wind vector). The model parameters include, besides the ones introduced in this section, the coefficients that link the angles $\alpha,\,\beta$ and their derivatives to the aerodynamic forces and moments, see \cite{Etkin}. This model is employed only after take-off. i.e. in the second operating mode described in section \ref{S:aerodynamics}. The initial state is derived by taking the state of the ground station at the time instant $t^*$ \eqref{E:t_star} (i.e. at take-off) and computing congruent values for the position, attitude, velocity and angular speed of the glider.
\section{Control design}\label{S:control}
As mentioned in the Introduction, we propose a decentralized control approach, where the controller of the ground station (respectively of the aircraft) compute the values of $u_{GS}$ (resp. $u_g$) according to local information. We assume that the two controllers are aware of whether the aircraft is on the slide (first operating mode in section \ref{S:aerodynamics}) or not (second operating mode). This information can be easily obtained with contact or proximity sensors installed on both the ground station and the glider.
\subsection{Ground station controller}\label{SS:ground_station_control}
\begin{figure}[hbt!]
 	\centerline{
 		\includegraphics[width=8cm,clip]{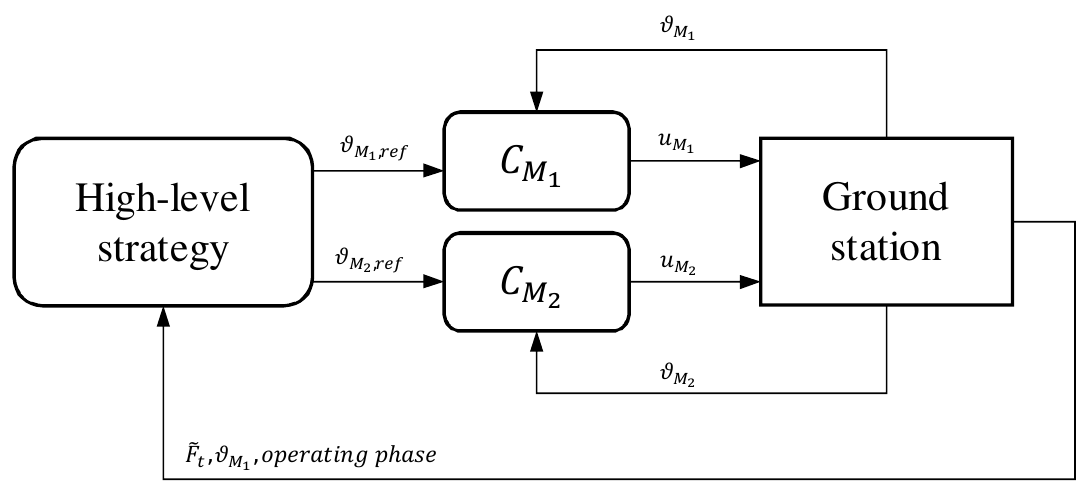}}
 	\caption{Controller for the ground station. The ``operating phase'' in the outer feedback path refers to whether the system is into the first or second operating mode; this is a boolean type of information that can be detected by means of e.g. a proximity switch.}\label{F:high_level_control_GS}
 \end{figure}

The controller for the ground station is hierarchial (see Fig. \ref{F:high_level_control_GS}): two low-level position control loops track the reference angular positions for motors $M_1$ and $M_2$, issued by a high-level strategy.
The low-level controllers are linear, designed using standard loop-shaping techniques \cite{SkPo05} since the ground station dynamics are essentially linear as long as the tether force is kept at zero, i.e. when the winch position is slightly larger than the glider position relative to the starting point (see \eqref{E:dyn_mod_first}, \eqref{E:dyn_mod_second} and \eqref{E:line_tension}). On the other hand, the high-level strategy depends on the operating mode. During the launching maneuver (first mode), a step of the reference position $\vartheta_{M_2,ref}$ for motor $M_2$ is commanded, in order to move the slide as fast as possible over a desired distance. During the motion, the slide reaches the take-off speed. At the same time, the reference position $\vartheta_{M_1,ref}$ for the winch motor $M_1$ is latched to the slide movement, with a slightly larger value. In this way, we avoid tensioning the line during the launching, which would result in a braking force on the slide, see equation \eqref{E:dyn_mod_first}. Thus, the high-level strategy in the first mode operates as follows:
\begin{equation}\label{E:high_level_first}
\begin{array}{rcl}
\vartheta_{M_1,\text{ref}}&=&\vartheta_{M_2}+\Delta\vartheta^I\\
\vartheta_{M_2,\text{ref}}&=&\dfrac{L}{r_{M_2}}
\end{array}
\end{equation}
where $L$ is the desired slide travel (limited by the length of the rails) and $\Delta\vartheta^I>0$ is a tuning parameter set to achieve a desired margin between the position of the two motors, in order to keep the tether slightly slack.

When the take-off speed has been reached and the aircraft detaches from the slide (second operating mode), the value of $\vartheta_{M_1,ref}$ is computed according to a different strategy, which aims to control the tether reeling in order to maintain a little tension at all time. In this way, the aircraft's dynamics are not affected significantly by the tether, and at the same time the amount of line sag is limited. In particular, we consider the measured line tension:
\begin{equation}\label{E:meas_force}
\tilde{F}_t=\|\vec{F}_t\|_2+v
\end{equation}
where $v$ is measurement noise, and we estimate the line elongation using an approximated, constant stiffness $\hat{K}$:
\begin{equation}\label{E:estim_pos}
\hat{P}=\dfrac{\tilde{F}_t}{\hat{K}}+\vartheta_{M_1}r_{M_1}.
\end{equation}
Then, we compute the reference position of the two low-level controllers as:
\begin{equation}\label{E:high_level_second}
\begin{array}{rcl}
\vartheta_{M_1,\text{ref}}&=&\dfrac{\hat{P}}{r_{M_1}}-\Delta\vartheta^{II}\\
\vartheta_{M_2,\text{ref}}&=&\dfrac{L}{r_{M_2}}
\end{array}
\end{equation}
where $\Delta\vartheta^{II}>0$ is a tuning parameter that corresponds to the desired line tensioning. The control strategy \eqref{E:high_level_second} is used throughout the flight and landing phases.

\subsection{Aircraft controller}\label{SS:flight_control}
\begin{figure}[hbt!]
 	\centerline{
 		\includegraphics[width=8cm,clip]{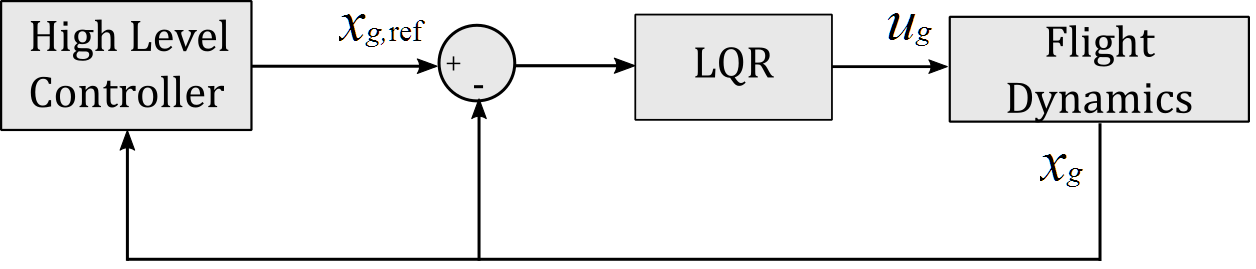}}
 	\caption{Controller scheme.}\label{D:Control}
 \end{figure}

The control system for the aircraft dynamics is hierarchical, too, as represented in Fig.~\ref{D:Control}. A low level regulator tracks a reference trim position for the glider's state during the flight. A second, high-level controller is used to compute such reference in order to control the flight path.

For the inner loop, we adopt a Linear Quadratic Regulator (LQR) designed considering the linearization of the system dynamics \eqref{E:F1} around a steady state $\vec{x}_{g,\textrm{trim}}$ and corresponding input $\vec{u}_{g,\textrm{trim}}$. The pair ($\vec{x}_{g,\textrm{trim}},\,\vec{u}_{g,\textrm{trim}}$) corresponds to a straight flight, constant altitude motion. The linearized dynamics are computed by neglecting the presence of the line, which is then an external disturbance from the point of view of the glider controller. Considering that the ground station control system modulates the line reeling in order to obtain a low tension, this approach is reasonable. Moreover, if the target flight pattern does not include sharp turns and sudden changes of altitude, a single linear LQR controller suffices for the whole cycle of take-off, flight and landing. We therefore decided not to adopt a gain-scheduling of the low-level controller.

Regarding the high-level controller, we set a sequence of target way-points in space, denoted as $[p_{i,X}^w,p_{i,Y}^w,p_{i,Z}^w]^T,\,i=1,\ldots,N$, that are used to compute reference altitude and heading for the low-level LQR. The way-points are chosen in order to achieve a roughly rectangular flight pattern, and the switching from one to the next way-point is based on a proximity condition.

\begin{figure}[!bht]
 	\centerline{
 		\includegraphics[width=8cm,clip]{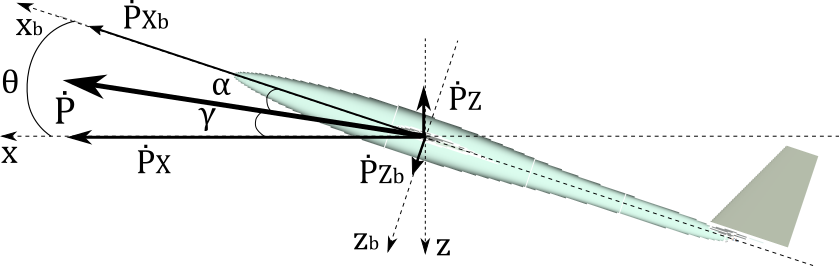}}
 	\caption{Longitudinal plan of motion and associated variables for the altitude flight controller.}\label{F:Alti}
 \end{figure}

For a given way-point, the high-level strategy issues two reference signals: one to control the altitude of the aircraft, and one to control its heading. The altitude controller computes a reference pitch rate $\omega_{Y_b,\textrm{ref}}$ on the basis of the measured  path angle $\gamma$, defined as:
 \begin{equation}\label{E:gamma}
\gamma\doteq\alpha-\theta,
\end{equation}
see  Fig.~\ref{F:Alti} for a graphical representation. 
A reference path angle $\gamma_{\textrm{ref}}$ is derived from the actual altitude and the altitude of the current target way-point:
\begin{equation}\label{E:gamma_ref}
\gamma_{\textrm{ref}}=\arctan\left(\frac{p_{i,Z}^w-p_Z}{p_{i,X}^w-p_X}\right).
\end{equation}
Then, the reference pitch rate given to the LQR is computed as:
\begin{equation}\label{E:pitch_rate_ref}
\begin{array}{rcl}
\omega_{Y_b,\textrm{ref}}&= & -k_\gamma (\gamma_{\textrm{ref}}-\gamma),
\end{array}
\end{equation}
where $k_\gamma$ is a constant gain chosen by the control designer. It can be shown that such a control approach provides stable closed-loop dynamics at least in the neighborhood of the chosen linearization point. Deriving \eqref{E:gamma} with respect to time and considering the presence of the low-level LQR that tracks a reference pitch rate, we obtain:
\begin{equation}\label{E:gamma_dot}
\begin{array}{rcl}
\dot{\gamma} & =& -\omega_{Y_b} + \dot{\alpha} \simeq -\omega_{Y_b,\textrm{ref}} + d,
\end{array}
\end{equation}
where $d$ is an additive term accounting for the rate of change of the angle of attack $\dot{\alpha}$, for the tracking error of the low-level controller on the pitch rate, and for eventual other sources of uncertainty (e.g. wind turbulence). This term can be reasonably assumed to be bounded in the considered flight conditions. Substituting \eqref{E:pitch_rate_ref} in (\ref{E:gamma_dot}) and applying the Laplace transform we obtain the following stable, first order system with the reference path angle and an external, bounded disturbance as inputs:
\begin{equation}\label{E:PathEquation}
\gamma(s) = \frac{k_\gamma}{s+k_\gamma}\gamma_{\textrm{ref}}(s) + \frac{1}{s+k_\gamma}d(s)
\end{equation}
where $s$ is the Laplace variable. The closed-loop system in \eqref{E:PathEquation} achieves tracking of the reference path angle and reduces the effects of disturbances when $k_\gamma$ is properly tuned. The value of $\gamma_{\textrm{ref}}$ in \eqref{E:gamma_ref} is saturated to avoid divergence problems when the glider is close to the target point and to avoid too large pitching commands. This control approach is particularly good for landing, where it adjusts the descent angle to aim at the origin.

\begin{figure}[hbt]
 	\centerline{
 		\includegraphics[width=8cm,clip]{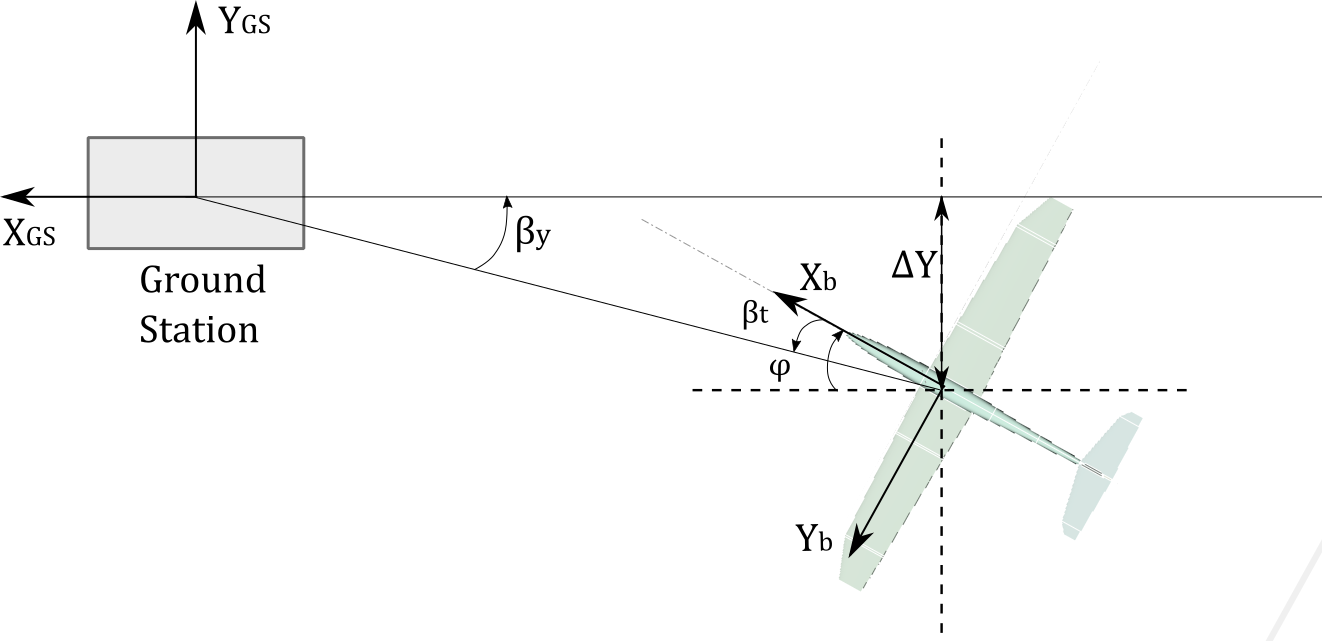}}
 	\caption{Lateral positioning analysis.}\label{D:lateral}
 \end{figure}

The second reference signal issued by the high level controller is for the heading of the glider. The reference heading needed to reach the current target point is computed as
\begin{equation}\label{E:control_heading}
\psi_{\textrm{ref}} = \arctan \left( \frac{p_{i,Y}^w - p_Y}{p_{i,X}^w- p_X} \right)
\end{equation}
where the four-quadrant arctangent is used. The LQR then tracks such a reference yaw angle. To obtain smooth transitions from one target point to the next, we filter the reference heading signal with a first order low-pass filter. Computing the yaw reference with \eqref{E:control_heading} is sufficient to control the heading during the flight. However, this approach does not consider the alignment of the glider with the orientation of the ground station, which is required to land with high accuracy. Hence, in the landing phase another strategy is used within the high-level controller. In particular, assuming without loss of generality that the last target point is the origin of the inertial system, we consider the angle $\beta_y$, defined as (see Fig.~\ref{D:lateral}):
\begin{equation}\label{E:beta_y}
\beta_y= \arctan\left(\dfrac{p_Y}{p_X}\right)= {} \beta_t+\psi
\end{equation}
where $\beta_t$ is the angle between the line projected on the ground, and the inertial $X$-axis, as shown in Fig.~\ref{D:lateral}. In a way similar to the altitude controller \eqref{E:gamma_ref}-\eqref{E:pitch_rate_ref}, we set a reference yaw rate as:
\begin{equation}\label{E:yaw_rate_ref}
\dot{\psi}_{\textrm{ref}} = k_\beta(\beta_{y_{\textrm{ref}}}-\beta_y)
\end{equation}
with $k_\beta$ being a design parameter.  With a procedure similar to the one used for the altitude controller, we can show that \eqref{E:yaw_rate_ref} gives rise to the following stable closed-loop dynamics
\begin{equation}\label{E:PathEquation2}
\beta_y(s) = \frac{k_\beta}{s+k_\beta}\beta_{y,\textrm{ref}}(s)\frac{1}{s+k_\beta}d_\beta(s)
\end{equation}%
where $d_\beta$ is a bounded disturbance term. In order to align the aircraft with the rails, we set $\beta_{y,\textrm{ref}}=0$ throughout the landing maneuver.
%
%
%
%
%
%
%

\section{Simulation results}\label{S:results}

\begin{figure}[hbt]
 	\centerline{
 		\includegraphics[width=7cm,clip]{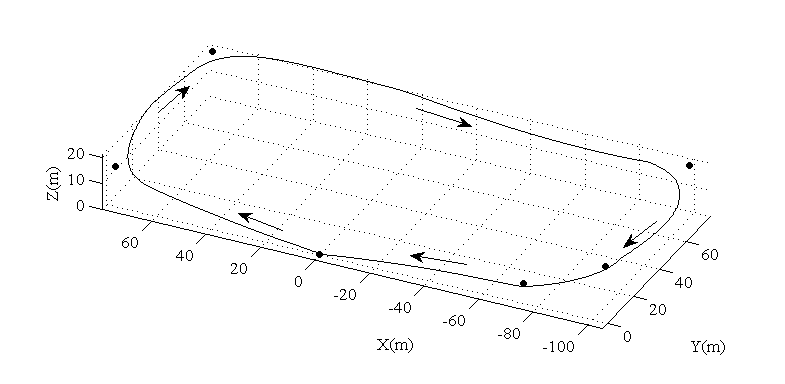}}
 	\caption{Simulation results with no wind. Three-dimensional illustration of the flight path with reference points.}\label{Path}
 \end{figure}

\begin{figure}[hbt]
 	\centerline{
 		\includegraphics[width=7cm,clip]{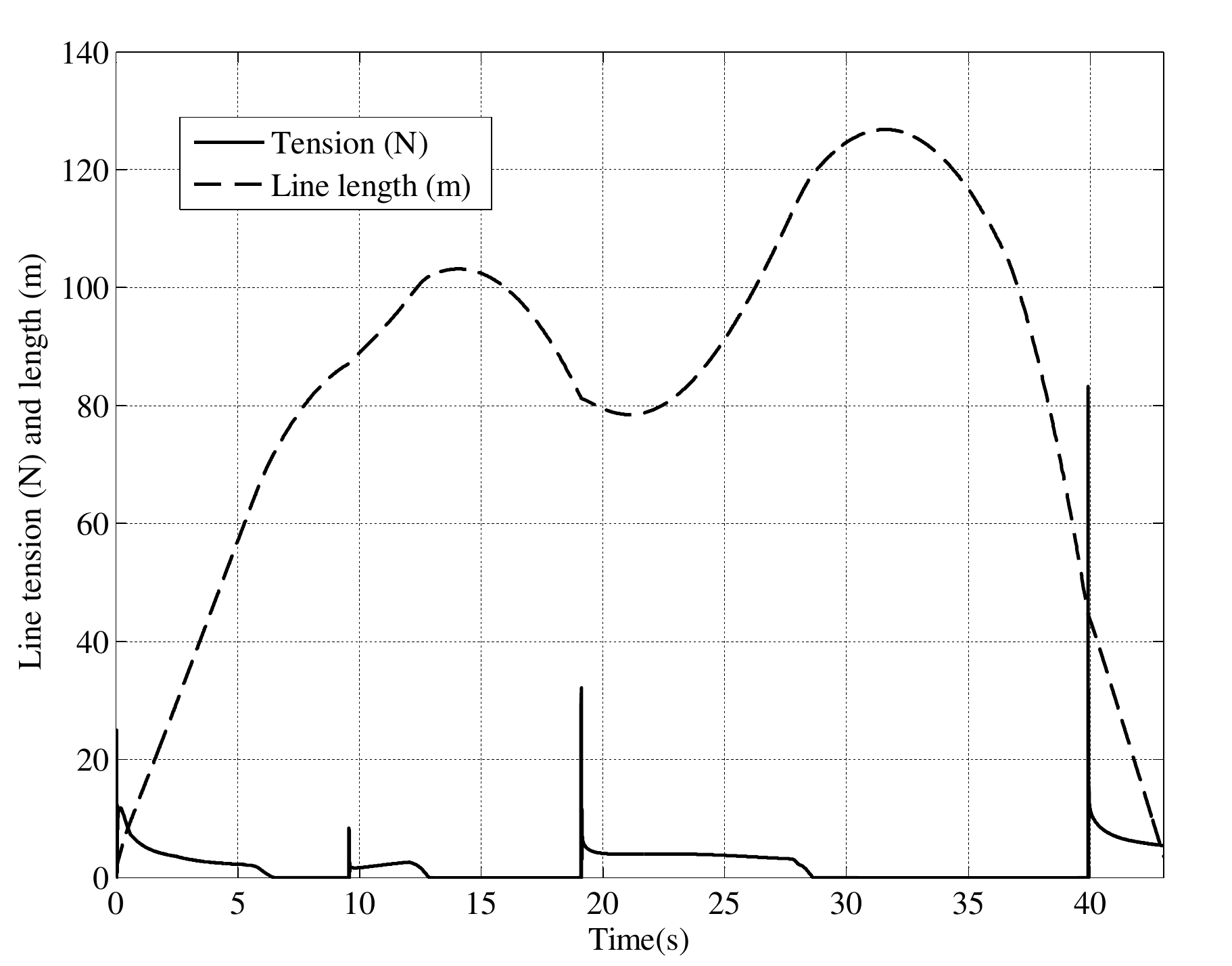}}
 	\caption{Simulation results with no wind. Tether tension (solid) and length (dashed) during flight and landing.}\label{F:line_tension_length}
 \end{figure}

We implemented the model and the control system in Matlab/Simulink. The main parameters of the glider are described in Table \ref{T:glider}.
\begin{table*}[h!]
	\caption{Model and control parameters employed in the numerical simulation study}
	\label{T:glider} \setlength{\tabcolsep}{10pt}
	\renewcommand{\arraystretch}{1.25}
	\centering
	\begin{tabular}{|l|ll|}
		\hline
\multicolumn{3}{|l|}{Aircraft and tether parameters}\\\hline
		Wingspan & 1.68&m  \\
		Aspect Ratio & 8.9& \\
		Wing loading & 3.8& Kg\,m$^{-2}$ \\
        Mass & 1.2 & Kg \\
		Onboard propulsion power & 180& W\\
        Tether Young modulus $E$ & 5.3$\,10^9$& Pa\\
        Tether breaking elongation $\varepsilon$ & 0.02&\\
        Tether drag coefficient $C_t$ & 1&\\
        Tether diameter $d_t$ & 0.002&m\\\hline\hline
\multicolumn{3}{|l|}{Ground station parameters}\\\hline
		$r_{M_1}$ & 0.1&m  \\
		$r_{M_2}$ & 0.1&m \\
		$J_{M_1}$ & 0.08& Kg m$^{2}$ \\
       $J_{M_2}$ & 0.01 & Kg m$^{2}$\\
		$\beta_{M_1}$ & 0.04 &Kg m$^2$ s$^{-1}$ \\
       $\beta_{M_2}$ & 0.01 & Kg m$^2$ s $^{-1}$\\
       $m_{s}$ & 9 & Kg\\
       $\beta_{s}$ & 0.6 & Kg s $^{-1}$\\
		\hline\hline
\multicolumn{3}{|l|}{Control parameters}\\\hline
$L$ & 5&m  \\
$\Delta\vartheta^I$ & $\pi$& rad  \\
$\Delta\vartheta^I$ & 0.1 &rad  \\
$\hat{K}$ & 500 & N m$^-1$  \\
$k_\gamma$ & 10&s$^-1$  \\
$k_\beta$ & 40&s$^-1$  \\
\hline
	\end{tabular}
	
\end{table*}
%
These parameters correspond to a model glider that we employ in experimental tests (currently ongoing) with the prototype shown in Fig. \ref{F:sys_pic}. Albeit the wing loading is much smaller than that of an aircraft designed for airborne wind energy, the presented results are easily scalable, and the same control approach can be used on a heavier aircraft. Indeed, a lighter aircraft makes the control problem more difficult, since there is a larger sensitivity to turbulence.
%
%
The full set of model and control parameters that we used is reported in the Appendix.
%
%
%
A typical simulated flight path, including the target points, is presented in Fig.~\ref{Path}. This path was obtained with no wind. It can be noted that the control system is able to control the glider while modulating the tether tension. The latter is shown in Fig. \ref{F:line_tension_length}. A low tension is kept throughout the flight, notwithstanding the pronounced changes in tether length, also shown in Fig. \ref{F:line_tension_length}, which matches closely the distance between the aircraft and the origin. The spikes of force in Fig. \ref{F:line_tension_length} correspond to the time instants when the tether is suddenly under tension from a slack condition (see \eqref{E:line_tension}). Even though such spikes are relatively small, they can be further mitigated by using a spring-damper system installed on the ground station. The LQR used for low-level controller of the aircraft was designed around an equilibrium point corresponding to a horizontal flight at $11$m$\,$s$^{-1}$. 
%

To assess the positioning precision achievable at landing with the proposed approach, we implemented a source of turbulence in the form of forward wind (along the negative $X$ direction) and wind gusts. Wind gusts are generated by a filtered white noise with zero mean. Their amplitude is a percentage of the nominal wind speed. These conditions are close from what one could expect for a take-off, assuming that it is always performed against the prevalent wind direction. We carried out four series of 100 simulated flights; each series had a different nominal wind speed and wind gusts with amplitude equal to $30\%$ of the nominal speed. The positioning precision at landing for each series is shown, together with the mean and standard deviation, in Table \ref{T:L}. The average and standard deviation of the landing speed complement these results.
\begin{figure}[hbt]
	\centerline{
		\begin{tabular}{cc}
			(a)&(b)\\			\includegraphics[width=0.25\textwidth]{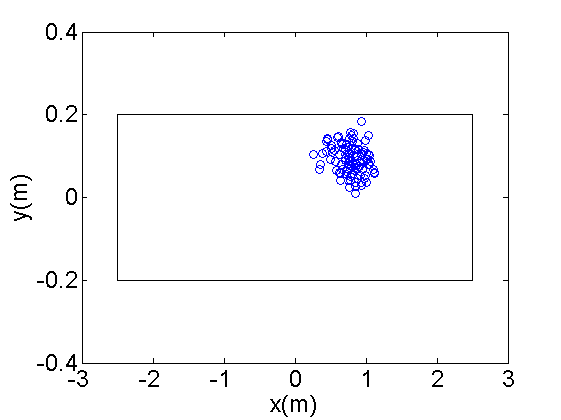}&
			\includegraphics[width=0.25\textwidth]{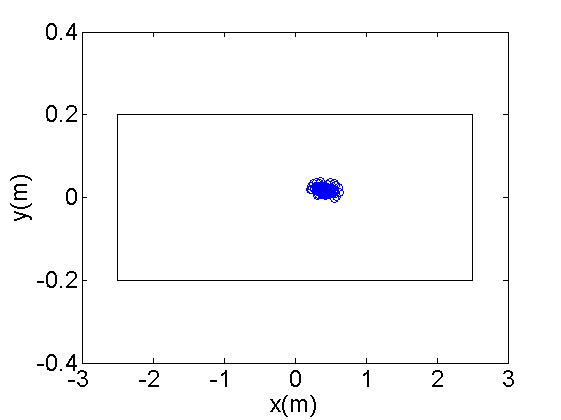}\\
			(c)&(d)\\
			\includegraphics[width=0.25\textwidth]{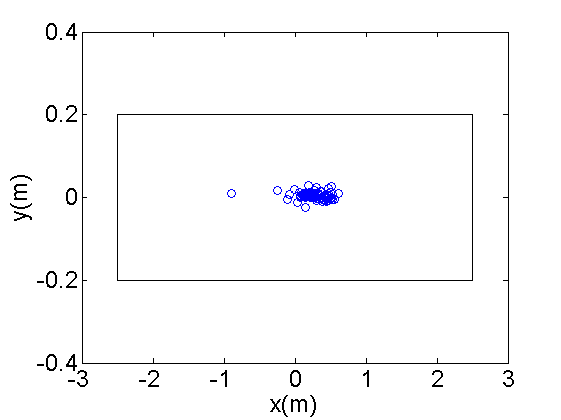}&
			\includegraphics[width=0.25\textwidth]{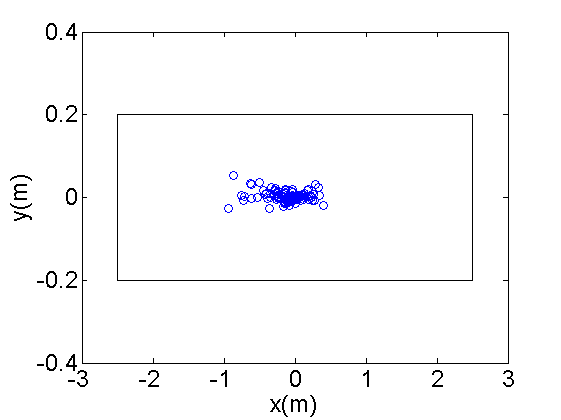}\\
		\end{tabular}
	} \protect\caption{Simulation results. Aircraft position at touch-down with nominal wind of (a) 1 m/s, (b) 2 m/s, (c) 3 m/s and (d) 4 m/s and uniformly distributed 3D wind disturbances in the range $\pm 30\%$ of the nominal wind. The solid rectangles in the plots correspond to the dimensions of the rails (note the different scales of the two axes).}\label{F:landing_accuracy}
\end{figure}
All the landing points are within the area spanned by the rails (0.4$\,$m $ \times 5\,$m). Figures \ref{F:landing_accuracy}(a)-(d) show the touch down points on the ground station with the origin being in the middle of the rails. The wind gusts do not affect much the landing precision, and the average touch-down point is pushed backward ($\bar{X}<0$m) with increasing nominal wind speed. It can be also noted that the precision of the lateral landing position improves with higher nominal wind. This is due to the fact that a higher front wind makes the aircraft speed relative to ground decrease, hence giving more time to align with the rails. All these effects are expected and indicate a good overall performance of the control system. The highest landing velocity measured during our tests is $13.7\textrm{m\,s}^{-1}$, while the highest average velocity is $12.1\textrm{m\,s}^{-1}$. These velocity values are obtained with aero-brakes (flaps) deployed for the whole landing phase. These devices efficiently reduce the velocity, which however remains quite large and should be reduced by more powerful drag generators or by adjusting the tension applied to the line during landing. 
This aspect will be investigated in more detail in the future, with both simulation and experiments.
%
%
\begin{table*}[hbt!]\center
	\caption{Landing precision and velocities for different wind conditions.\label{T:L}}
	\centering
	\begin{tabular}{|l|c|c|c|c|c|c|}
		\hline
		Condition &  X mean (m) & $\sigma_X$ (m) & Y mean (m) & $\sigma_Y$ (m) & Velocity (m.s$^{-1}$) & $\sigma_V$ (m.s$^{-1}$)\\
		\hline
		Wind gust ($1$m/s)& $0.78$ & $0.20$ & $0.091$ & $0.037$ & $12.1$ & $1.32$\\
		$2$m/s Wind + $30\%$ wind gusts & $0.41$ & $0.10$ & $0.019$ & $0.008$ & $11.0$ & $1.17$\\
		$4$m/s Wind + $30\%$ wind gusts & $0.28$ & $0.20$ & $0.004$ & $0.008$& $9.54$ & $1.17$\\
		$6$m/s Wind + $30\%$ wind gusts & $-0.11$ & $0.26$ & $0.003$ & $0.013$ & $7.04$ & $0.74$\\
		\hline
	\end{tabular}
\end{table*}

\section*{Appendix\\Model and control parameters}

We list here the full set of parameters and the equations we used to determine the aerodynamic coefficients of the aircraft's model.
The aerodynamic forces are functions of the state  and the euler angles. They can be expressed as a sum of independent non-dimensional coefficients. Etkin \cite{Etkin} gives the main coefficients in our case, assuming low Mach number and neglecting the body contribution:
\small
\begin{align*}
F_{a,X_b} = {} & QS_w \left[ C_{\textrm{d}}(\alpha_w,\alpha_t) + C_{\textrm{d} u_f}  u_{f}\right]\\
F_{a,Y_b}= {} & QS_w \left[C_{\textrm{Y}\beta} \beta + C_{\textrm{Y}\hat{p}} \hat{p} + C_{\textrm{Y}\hat{r}} \hat{r} + C_{\textrm{Y}u_r} u_{r} \right]\\
F_{a,Z_b} = {} & QS_w \left[ C_L(\alpha_w,\alpha_t) + C_{L\hat{q}} \hat{q} + C_{L\hat{\dot{\alpha}}} \hat{\dot{\alpha}} + C_{\textrm{L}u_e} u_{e} + C_{\textrm{L} u_{f}} u_{f} \right]\\
M_{X_b}= {} & QS_w \left[ C_{l\beta} \beta + C_{l\hat{p}} \hat{p} + C_{l\hat{r}} \hat{r} + C_{\textrm{l}u_a} u_a + C_{\textrm{l} u_r} u_r\right]\\
M_{Y_b}= {} & QS_w \left[ C_{\textrm{m0}} + C_{\textrm{m}\alpha} \alpha + C_{\textrm{m}\hat{\dot{\alpha}}} \hat{\dot{\alpha}} + C_{\textrm{m}\hat{q}} \hat{q} + C_{\textrm{m} u_e}  u_e + C_{\textrm{m}u_f}  u_{f}\right]\\
M_{Z_b}= {} & QS_w \left[ C_{\textrm{n}\beta} + Cn_{\hat{p}} \hat{p} + Cn_{\hat{r}} \hat{r}+ C_{\textrm{l} u_a} u_a + C_{\textrm{l} u_r} u_r \right]
\end{align*}
\normalsize
Where $Q=0.5\rho W_a^2$ is the dynamical pressure, $\hat{p}=\frac{\omega_{X_b} b}{2W_a}$, $\hat{q}=\frac{\omega_{Y_b} c}{2W_a}$, $\hat{r}=\frac{\omega_{Z_b} b}{2W_a}$ are the non-dimensional rotation rates and $\hat{\dot{\alpha}}=\frac{c \dot{\alpha}}{2W_a}$ is the non-dimensional $\dot{\alpha}$.
Note that to account for the aero-breaks forces, the flaps coefficients and inputs are simply replaced by the ones of the aero-breaks, since they use the same actuators in combination with ailerons.
To determine the aerodynamic coefficients, we used XFLR5, a software based on Xfoil \cite{XFLR5}. This software was developed especially for low Reynolds number analysis and model airplane design. When a coefficient couldn't be determine using XFLR5, an empirical formula from \cite{Etkin} was used. The aerodynamic derivatives of the inputs were also estimated from XFLR5, except for the thrust for which we assumed a linear forward force ranging from 0 to $F_{\textrm{Thrust}}$.

The glider we use in our experiments was modeled in XFLR5 using the geometric parameters given in Table \ref{T:AeroGeo}. The airfoil shape was guessed from a list of well-know airfoils for model gliders. For the main wing, a S7012 airfoil is used while for the tail a symmetric airfoil, NACA 009, is used.

The aerodynamic coefficients as given by XFLR5 are resumed in Table \ref{T:Aero coef} .

The remaining coefficients are deduced derivatives defined in Table \ref{T:Aero coef}. For convenience, $\frac{dC_{Lw}}{d\alpha}$, $\frac{dC_L}{d\alpha_f}$ and $\frac{dC_Y}{d\alpha_t}$ are replaced by $a_w$, $a_f$ and $a_t$.

\begin{equation}\nonumber
\begin{array}{rcl}
\multicolumn{3}{l}{\textbf{Drag coefficients:}}\\
C_d(\alpha_w,\alpha_t) &=&C_d(\alpha_w)+C_{dt}(\alpha_t)\\
\multicolumn{3}{l}{\text{Wing Drag coefficient}} \\
    C_d(\alpha_w)&=& {} C_{d0} + \frac{C_{Lw}(\alpha_w)^2}{\pi e AR}\\
\multicolumn{3}{l}{\text{Tail Drag coefficient}} \\
    C_{dt}(\alpha_t)&=& {} \frac{S_t}{S_w} \left( C_{d0}+\frac{C_{Lt(\alpha_t)}^2}{\pi e AR} \right)\\
    \\
    \multicolumn{3}{l}{\textbf{Side Force Coefficients:}}\\
	\multicolumn{3}{l}{\text{Roll-rate coefficient}} \\
	C_{Y\hat{p}}&=& {} -2 a_f \frac{S_f z_f}{S_w b}\\
	\multicolumn{3}{l}{\text{Yaw-rate coefficient}} \\
	C_{Y\hat{r}}&=&  {} 2 a_f \frac{S_f l_f}{S_w b}\\
	\\
\multicolumn{3}{l}{\textbf{Lift coefficients:}}\\
C_L(\alpha_w,\alpha_t) & =& C_{Lw}(\alpha_w) + C_{Lt}(\alpha_w, \alpha_t)\\
\multicolumn{3}{l}{\text{Wing Lift coefficient}}\\
	C_{Lw}(\alpha_w)&=& {} \alpha_w a_w + C_{L0} \\
\multicolumn{3}{l}{\text{Tail Lift coefficient}} \\
	C_{Lt}(\alpha_w, \alpha_t) &=& {} a_t(\alpha_w (1-\epsilon) + \alpha_t) \frac{S_t}{S_w}\\
\multicolumn{3}{l}{\text{AoA general derivative, wing and tail}}\\
	C_{La}&=&a_w \left(1+\frac{a_t S_t}{a_w S_w}(1-\epsilon) \right)\\
\multicolumn{3}{l}{\text{Pitch-rate coefficient}}\\
	C_{L\hat{q}}&=& a_w (0.5 + 2 \|x_{ac}-x_{cg}\|)\\&&-2 a_t \eta \frac{S_t}{S_w} \left(\frac{l_t}{c}-x_{cg}\right)\\
\multicolumn{3}{l}{\text {AoA rate coefficient}} \\
	C_{L\dot{\alpha}}&=& 2 a_t V_h \epsilon\\
	\\
\multicolumn{3}{l}{\textbf{Rolling moment coefficients:}} \\
\multicolumn{3}{l}{\text{Side slip coefficient}} \\
	C_{l\beta}&=& -2 a_f \frac{S_f z_f}{S_w b}\\
\multicolumn{3}{l}{\text{Yaw-rate coefficient (Mainly from tail)}} \\
	C_{\hat{r}t}&=& 2 a_f \frac{S_f l_f}{S_w b} \eta\\
\multicolumn{3}{l}{\text{Roll-rate coefficient (Mainly from wing)}} \\
	C_{\hat{p}}&= & -0.5 C_{La} \frac{1+3\lambda}{12(1+\lambda)}\\
	\\
\multicolumn{3}{l}{\textbf{Pitching moment coefficients:}}\\
	Cm_{\alpha} & = &C_{mw}+C_{mt}\\
\multicolumn{3}{l}{\text{Wing pitch moment coeff.}}\\
	C_{mw}&= & C_{Lw} (x_{cg} -x_{ac}) + C_{mf} \cdot d_f\\
\multicolumn{3}{l}{\text{Tail pitch moment coeff.}} \\
	C_{mt}&= & - C_{Lt} (\frac{l_t}{c}-(x_{cg}-x_{ac}))\\
\multicolumn{3}{l}{\text{Tail pitch-rate moment coeff.}} \\
	C_{m\hat{q}t} &=& -2 a_t V_h \frac{l_t}{c}\\
\end{array}
\end{equation}
\begin{equation}\nonumber
\begin{array}{rcl}
\multicolumn{3}{l}{\text{Wing pitch-rate moment coeff.}} \\
	C_{m\hat{q}w} &= & -2 C_{La} (x_{cg}-0.5)^2\\
\multicolumn{3}{l}{\dot{\alpha}\text{ moment coeff.}} \\
	C_{m\hat{\dot{\alpha}}}&=& -2 a_t \epsilon V_h \frac{l_t}{c}\\
\multicolumn{3}{l}{\text{Elevator moment coeff.}} \\
	C_{mu_e} &=& -C_{Lt} \left(\frac{l_t}{c}-(x_{cg}-x_{ac})\frac{S_t}{S_w}\right)\\
	\\
\multicolumn{3}{l}{\textbf{Yawing moment coefficients:}} \notag \\
\multicolumn{3}{l}{\text{Roll-rate moment coeff.}} \\
	Cn_{\hat{p}}&=& 2 C_{Yp} \frac{l_f}{b}\\
\multicolumn{3}{l}{\text{Yaw-rate moment coeff.}} \\
	Cn_{\hat{r}} &=& - 2 a_f V_v \eta \frac{l_f}{b}\\
\multicolumn{3}{l}{\text{Rudder moment coeff.}} \\
	Cn_{u_r} &=&C_{Yr} \frac{S_w}{S_f} V_v
\end{array}
\end{equation}

\begin{table*}[h!]
	\caption{Geometry and mass constants for the employed glider.}
	\label{T:AeroGeo}\setlength{\tabcolsep}{10pt}
	\renewcommand{\arraystretch}{1}
	\centering
	\begin{tabular}{|l|c|}
		\hline
		Geometry &  Value \\
		\hline
		Wing Span & $b=1.68$m\\
		Wing Surface Area & $S_=0.317 \textrm{m}^{2} $ \\
		Horizontal Tail Surface Area & $S_t=0.059 \textrm{m}^{2} $ \\
		Fin Surface Area & $S_f=0.024 \textrm{m}^{2} $ \\
		Mean Aerodynamic Chord (MAC)& $c=0.194$m\\
		Horizontal Tail MAC & $c_t=0.115$m\\
		Horizontal Tail, horizontal level arm & $l_t=0.658$m\\
		Horizontal Tail, vertical level arm & $z_t=0$m\\
		Fin, horizontal level arm & $l_f=0.657$m\\
		Fin, vertical level arm & $z_f=0.1$m\\
		Horizontal Tail Volume & $V_h=\frac{S_t l_t}{S_w c}=0.63$\\
		Fin Volume & $V_f=\frac{S_f l_f}{S_w c}=0.0296$\\
		Wing Aspect ratio & $AR=8.89$\\
		Horizontal Tail, Aspect ratio & $AR_t=4.94$\\
		Oswald coefficient for the wing & $e=0.8$\\
		Tapered ratio & $\lambda=2.35$\\
		Aerodynamic center (relative to MAC) & $x_{ac}=0.25\cdot c$\\
		Center of mass (relative to MAC) & $x_{cg}=0.23\cdot c$\\
		\hline
		Mass parameters & Value\\
		\hline
		Mass & $1.2$Kg\\
		Inertia $I_{\textrm{xy}}$ & $-0.00275\textrm{Kg.m}^2$\\
		Inertia $I_{\textrm{xx}}$ & $0.0576\textrm{Kg.m}^2$\\
		Inertia $I_{\textrm{yy}}$ & $0.103\textrm{Kg.m}^2$\\
		Inertia $I_{\textrm{zz}}$ & $0.1598\textrm{Kg.m}^2$\\
		\hline
		Other & Value\\
		\hline
		Oswald coefficient for the wing & $e=0.8$\\
		Air velocity ratio between wing and Tail& $\eta=0.9$\\
		Down-wash acting on the Tail & $\epsilon = \frac{a_w}{\pi AR e}=0.203$\\\hline
	\end{tabular}
\end{table*}

\begin{table*}[h!]
	\caption{Aerodynamic derivatives for the employed glider computed by XFLR5.}
	\label{T:Aero coef}\setlength{\tabcolsep}{10pt}
	\renewcommand{\arraystretch}{1}
	\centering
	\begin{tabular}{|l|c|}
		\hline
		Derivatives &  Value \\
		\hline
		$\frac{dC_{Lw}}{d\alpha}$ Wing only & $4.81 rad^{-1} $ \\
		$\frac{dC_Y}{d\alpha_t}$ Tail only & $4.07 rad^{-1}$ \\
		$\frac{dC_L}{d\alpha_f}$ Fin only & $3.29 rad^{-1}$ \\
		$C_{Y\beta}$ wing side force & $-0.17 rad^{-1}$\\
		$C_{lw}$ wing induced roll moment & $-0.0055 rad^{-1}$\\
		$C_{n\beta}$ side slip induced yaw moment & $0.0745 rad^{-1}$\\
		\hline
		Constants & Value\\
		\hline
		$C_{L0} (\alpha_w=0) $ Wing & 0.139\\
		$C_{L0} (\alpha_t=0) $ Tail & 0\\
		$C_{d0} (\alpha_w=0)$ Wing & 0.0115\\
		$C_{d0} (\alpha_t=0)$ Tail & 0.0145\\
		\hline
		Input derivatives & Value\\
		\hline
		$C_{\textrm{l}u_a}$, Aileron & $0.0061/\degree$\\
		$C_{\textrm{n} u_a}$, Aileron & $-0.00068/\degree$\\
		$C_{\textrm{L} u_e}$, Elevator & $0.048/\degree$\\
		$C_{\textrm{Y} u_r}$, Rudder & $0.048/\degree$\\
		$C_{\textrm{L} u_f}$, Flap	& $0.015/\degree$\\
		$C_{\textrm{d} u_f}$, Flap & $0.00066/\degree$\\
		$C_{\textrm{m} u_f}$, Flap & $-0.003/\degree$\\
		$C_{\textrm{L} u_{ab}}$, Aero-Break  & $-0.015/\degree$\\
		$C_{\textrm{d} u_{ab}}$, Aero-Break  & $0.008/\degree$\\
		$C_{\textrm{m} u_{ab}}$, Aero-Break  & $-0.001/\degree$\\\hline
	\end{tabular}
\end{table*}

\bibliographystyle{plain}

\begin{thebibliography}{10}

\bibitem{ampyx}
Ampyx Power website, http://www.ampyxpower.com/.

\bibitem{XFLR5}
Xflr5 guidelines, 2013.

\bibitem{AWEbook}
U.~Ahrens, M.~Diehl, and R.~Schmehl, editors.
\newblock {\em Airborne Wind Energy}.
\newblock Green Energy and Technology. Springer-Verlag Berlin, 2014.

\bibitem{ch5-Arch15}
C.L. Archer.
\newblock {\em Airborne Wind Energy}, chapter 5. An Introduction to Meteorology
  for Airborne Wind Energy, page~81.
\newblock Green Energy and Technology. Springer-Verlag, Berlin, 2014.

\bibitem{ArCa09}
C.L. Archer and K.~Caldeira.
\newblock Global assessment of high-altitude wind power.
\newblock {\em Energies}, 2(2):307--319, 2009.

\bibitem{BaOc12}
J.~H. Baayen and W.~J. Ockels.
\newblock Tracking control with adaption of kites.
\newblock {\em IET Control Theory and Applications}, 6(2):182--191, 2012.

\bibitem{Bont10}
Eelke Bontekoe.
\newblock Up! - how to launch and retrieve a tethered aircraft.
\newblock Master's thesis, TU Delft, August 2010.
\newblock Accessed in August 2015 at http://repository.tudelft.nl/.

\bibitem{ch24-BRKGS14}
A.~Bormann, M.~Ranneberg, P.~K\"{o}vesdi, C.~Gebhardt, and S.~Skutnik.
\newblock {\em Airborne Wind Energy}, chapter 24. Development of a Three-Line
  Ground-Actuated Airborne Wind Energy Converter, page 427.
\newblock Green Energy and Technology. Springer-Verlag, Berlin, 2014.

\bibitem{ch17-BSTR14}
A.~Bosch, R.~Schmehl, P.~Tiso, and D.~Rixen.
\newblock {\em Airborne Wind Energy}, chapter 17. Nonlinear Aeroelasticity,
  Flight Dynamics and Control of a Flexible Membrane Traction Kite, page 307.
\newblock Green Energy and Technology. Springer-Verlag, Berlin, 2014.

\bibitem{BSTR14}
A.~Bosch, R.~Schmehl, P.~Tiso, and D.~Rixen.
\newblock Dynamic nonlinear aeroelastic model of a kite for power generation.
\newblock {\em AIAA Journal of Guidance, Control and Dynamics},
  37(5):1426--1436, 2014.

\bibitem{ch16-BrSO14}
J.~Breukels, R.~Schmehl, and W.~Ockels.
\newblock {\em Airborne Wind Energy}, chapter 16. Aeroelastic Simulation of
  Flexible Membrane Wings based on Multibody System Dynamics, page 287.
\newblock Green Energy and Technology. Springer-Verlag, Berlin, 2014.

\bibitem{CaFM09c}
M.~Canale, L.~Fagiano, and M.~Milanese.
\newblock High altitude wind energy generation using controlled power kites.
\newblock {\em IEEE Transactions on Control Systems Technology}, 18(2):279
  --293, mar. 2010.

\bibitem{ErSt14}
Michael Erhard and Hans Strauch.
\newblock Flight control of tethered kites in autonomous pumping cycles for
  airborne wind energy.
\newblock {\em {Control Engineering Practice, submitted. preprint available on
  arXiv:1409.3083}}.

\bibitem{ErSt12}
Michael Erhard and Hans Strauch.
\newblock Control of towing kites for seagoing vessels.
\newblock {\em IEEE Transactions on Control Systems Technology}, 21(5):1629 --
  1640, 2013.

\bibitem{Etkin}
B.~Etkin.
\newblock {\em Dynamics of Atmospheric Flight}.
\newblock N.Y. : Dover Publication, 1972.

\bibitem{ch18-GoLu14}
R.~H.~Luchsinger F.~Gohl.
\newblock {\em Airborne Wind Energy}, chapter 18. Simulation Based Wing Design
  for Kite Power, page 325.
\newblock Green Energy and Technology. Springer-Verlag, Berlin, 2014.

\bibitem{FHBK14}
L.~Fagiano, K.~Huynh, B.~Bamieh, and M.~Khammash.
\newblock On sensor fusion for airborne wind energy systems.
\newblock {\em IEEE Transactions on Control Systems Technology},
  22(3):930--943, 2014.

\bibitem{FaMa15}
L.~Fagiano and T.~Marks.
\newblock Design of a small-scale prototype for research in airborne wind
  energy.
\newblock {\em IEEE/ASME Transactions on Mechatronics}, 20(1):166--177, 2014.

\bibitem{FaMi12}
L.~Fagiano and M.~Milanese.
\newblock Airborne wind energy: an overview.
\newblock In {\em American Control Conference 2012}, pages 3132--3143,
  Montreal, Canada, 2012.

\bibitem{FaMP09}
L.~Fagiano, M.~Milanese, and D.~Piga.
\newblock High-altitude wind power generation.
\newblock {\em IEEE Transactions on Energy Conversion}, 25(1):168 --180, mar.
  2010.

\bibitem{FZMK14}
L.~Fagiano, A.U. Zgraggen, M.~Morari, and M.~Khammash.
\newblock Automatic crosswind flight of tethered wings for airborne wind
  energy: modeling, control design and experimental results.
\newblock {\em IEEE Transactions on Control Systems Technology},
  22(4):1433--1447, 2014.

\bibitem{FeSc14}
U.~Fechner and R.~Schmehl.
\newblock Feed-forward control of kite power systems.
\newblock {\em Journal of Physics: Conference Series}, 524:012081, 2014.

\bibitem{ch20-ZiHa15}
F.~Fritz.
\newblock {\em Airborne Wind Energy}, chapter 20. Application of an Automated
  Kite System for Ship Propulsion and Power Generation, page 359.
\newblock Green Energy and Technology. Springer-Verlag, Berlin, 2014.

\bibitem{HoDi07}
B.~Houska and M.~Diehl.
\newblock Optimal control for power generating kites.
\newblock In {\em 9$^{th}$ European Control Conference}, pages 3560--3567, Kos,
  GR, 2007.

\bibitem{ch19-LRBLJP14}
R.~Leloup, K.~Roncin, G.~Bles, J.B. Leroux, C.~Jochum, and Y.~Parlier.
\newblock {\em Airborne Wind Energy}, chapter 19. Estimation of the
  Lift-to-Drag Ratio Using the Lifting Line Method: Application to a Leading
  Edge Inflatable Kite, page 339.
\newblock Green Energy and Technology. Springer-Verlag, Berlin, 2014.

\bibitem{ch28-Vand14}
D.~Vander Lind.
\newblock {\em Airborne Wind Energy}, chapter 28. Analysis and Flight Test
  Validation of High Performance Airborne Wind Turbines, page 473.
\newblock Green Energy and Technology. Springer-Verlag, Berlin, 2014.

\bibitem{ch21-MiTM14}
M.~Milanese, F.~Taddei, and S.~Milanese.
\newblock {\em Airborne Wind Energy}, chapter 21. Design and Testing of a 60 kW
  Yo-Yo Airborne Wind Energy Generator, page 373.
\newblock Green Energy and Technology. Springer-Verlag, Berlin, 2014.

\bibitem{ch26-RuSo14}
R.~Ruiterkamp and S\"{o}ren Sieberling.
\newblock {\em Airborne Wind Energy}, chapter 26. Description and Preliminary
  Test Results of a Six Degrees of Freedom Rigid Wing Pumping System, page 443.
\newblock Green Energy and Technology. Springer-Verlag, Berlin, 2014.

\bibitem{SkPo05}
S.~Skogestad and I.~Postlethwaite.
\newblock {\em Multivariable Feedback Control. 2$^{nd}$ edition}.
\newblock Wiley, 2005.

\bibitem{SHVDD15}
J.~Stuyts, G.~Horn, W.~Vandermeulen, J.~Driesen, and M.~Diehl.
\newblock Effect of the electrical energy conversion on optimal cycles for
  pumping airborne wind energy.
\newblock {\em IEEE Transactions on Sustainable Energy}, 6(1):2--10, 2015.

\bibitem{ch23-vdVlPS14}
R.~van~der Vlugt, J.~Peschel, and R.~Schmehl.
\newblock {\em Airborne Wind Energy}, chapter 23. Design and Experimental
  Characterization of a Pumping Kite Power System, page 403.
\newblock Green Energy and Technology. Springer-Verlag, Berlin, 2014.

\bibitem{ch30-VeGR14}
C.~Vermillion, B.~Glass, and A.~Rein.
\newblock {\em Airborne Wind Energy}, chapter 30. Lighter-Than-Air Wind Energy
  Systems, page 501.
\newblock Green Energy and Technology. Springer-Verlag, Berlin, 2014.

\bibitem{ZaGD13}
M.~Zanon, S.~Gros, and M.~Diehl.
\newblock Rotational start-up of tethered airplanes based on nonlinear mpc and
  mhe.
\newblock In {\em European Control Conference (ECC) 2013}, pages 1023--1028,
  Zuerich, Switzerland, July 2013, 2013.

\bibitem{ch12-ZaGD14}
M.~Zanon, S.~Gros, and M~Diehl.
\newblock {\em Airborne Wind Energy}, chapter 12. Model Predictive Control of
  Rigid-Airfoil Airborne Wind Energy Systems, page 219.
\newblock Green Energy and Technology. Springer-Verlag, Berlin, 2014.

\bibitem{ZaFM15}
A.U. Zgraggen, L.~Fagiano, and M.~Morari.
\newblock Real-time optimization and adaptation of the crosswind flight of
  tethered wings for airborne wind energy.
\newblock {\em IEEE Transactions on Control Systems Technology},
  23(2):434--448, 2015.

\bibitem{ch7-ZiHa15}
U.~Zillmann and S.~Hach.
\newblock {\em Airborne Wind Energy}, chapter 7. Financing Strategies for
  Airborne Wind Energy, page 117.
\newblock Green Energy and Technology. Springer-Verlag, Berlin, 2014.

\end{thebibliography}

\end{document}